\documentclass[aps,prd,floats,floatfix,twocolumn,showpacs]{revtex4}

\usepackage{amsmath}

\begin{document}

\title{The Liu-Yau mass as a quasi-local  energy in  general
relativity}

\author{Niall \'O Murchadha}
\email{niall@ucc.ie}
\affiliation{Physics Department, University College Cork, Ireland.}

\date{\today}

\begin{abstract}
A quasi-local mass has been a long sought after quantity in general 
relativity. A recent candidate has been the Liu-Yau mass. One
can show that the Liu-Yau mass of any two-surface is the maximum of
the Brown-York energy for that two-surface. This means that it has 
significant disadvantages as a mass.  It is much better interpreted as an energy and I will show one way of doing so. 
The Liu-Yau mass is especially interesting in spherical
geometries, where mass and energy are indistinguishable.  For a spherical two-surface, it equals the minimum of the amount
of energy at rest that one needs to put inside the two-surface to generate the given surface geometry. Thus it gives interesting information about the
interior, something no other mass or energy function does. 
\end{abstract}






\maketitle




General relativity is a gauge theory like
electromagnetism, and the metric $g_{\mu\nu}$ is the analogue of the
electromagnetic potential. For example the gauge freedom in electromagnetism
is represented by $A_{\mu} \rightarrow A_{\mu} +  \phi_{,\mu}$ while
infinitesmal coordinate transformations in  general relativity 
cause $g_{\mu\nu} \rightarrow g_{\mu\nu} + \lambda_{\mu,\nu} +
\lambda_{\nu,\mu}$. Therefore in gravity one might expect the
gauge-independent variables to be the first derivatives of the
potentials, as in either electromagnetism or Yang-Mills theory. However, in
general relativity  the gauge freedom, coordinate transformations, is such
that at any one point one can set the metric equal to the Minkowski metric and
set all the first derivatives of the metric at that point to zero. 

One consequence of this is that there exists in gravity no analogue of
$(\vec E \cdot \vec D + \vec B \cdot \vec H)/2$, the local electromagnetic energy density. Any attempt to
localize the gravitational energy density must involve a particular choice of
gauge.

Nevertheless, in gravity there does exist the concept of bulk energy,
expressed most clearly in the total, ADM, energy-momentum\cite{ADM}, with
\begin{equation}
E_{ADM} = \frac{1} {16\pi}\int_{\infty}(g_{ij,j} - g_{jj,i})dS_i
\end{equation}
 and
\begin{equation}
P^i_{ADM} = \frac{1}{8\pi}\int_{\infty} \pi^{ij}dS_j.
\end{equation}
These are defined on any asymptotically flat spacelike slice with
appropriate asymptotically cartesian coordinates, where $g_{ij}$ is the three
metric and $\pi^{ij}$ is its conjugate momentum. We can make a Lorentz boost,
find another suitable spacelike slice, and discover that the ADM
energy-momentum transforms like a Lorentz 4-vector  \cite{EOM}. Finally, one has
$M_{ADM} =
\sqrt{E^2_{ADM} - P^2_{ADM}}$. This is a boost invariant quantity which
mimics the Kepler mass at infinity. 

The existence of the $ADM$ energy-momentum
has led people to seek a quasi-local energy or mass, an object which is
somehow analogous to the surface-integral expression which relates the
electric flux-density
$\vec D$ to the total charge in the interior, or as in Newtonian gravity,
where the surface integral of the gradient of the potential on any closed two-surface measures the
interior mass content. Given a two-surface, one looks for an integral which
somehow quantifies the energy (or mass) contained within it. A very comprehensive review is given by Szabados \cite{sz}.

Brown and York \cite{BY} proposed a quasi-local energy-momentum vector in
this spirit. One starts off with a four-manifold (satisfying the Einstein
equations, with or without sources), one then chooses a spacelike
three-surface in this manifold, and finally one looks at a two-surface, $^{(2)}B$,
in the three-manifold. They assume that this 2-manifold has positive Gauss curvature so that the 2-surface can be isometrically embedded in flat 3-space. The Brown-York quasi-local energy density of this surface is
defined to be
\begin{equation}
e_{BY} = \frac{1} {8\pi} (k_0 - k), \label{BYe}
\end{equation}
where $k$ is the mean curvature of the two-surface as an
embedded surface in the three-space and $k_0$ is the mean curvature of the
isometric two-surface embedded in flat three-space. I define the mean
curvature of a round two-sphere of radius $R$ in flat space to be
$+2/R$.

The associated surface momentum density, as given by Brown, Lau, and York \cite{BLY} is
\begin{equation}
p_{iBLY} = \frac{1}{8\pi} n_j \pi_i^j /\sqrt{g},\label{BLYp}
\end{equation}
where $n^i$ is the normal of the 2-surface as embedded in the 3-space.

To define the total Brown-York energy, all one has to do is integrate the energy density over the 2-surface to give
\begin{equation}
E_{BY} = \frac{1} {8\pi}\int (k_0 - k)dA. \label{BYE}
\end{equation}
To define the total Brown-Lau-York momentum is somewhat trickier. We need to find three 3-vectors on the 2-surface, analogues of the translational Killing vectors of flat space, call them $\xi^i_{(A)}$, where $A$ runs through 1, 2, 3. We then define the total momentum as
\begin{equation}
P^{(A)}_{BLY} = \frac{1}{8\pi}\int \xi^i_{(A)} n_j \pi_i^j /\sqrt{g}dA.\label{BLYP}
\end{equation}

One way of choosing these approximate Killing vectors is to use the isometric embedding. This gives a flat Cartesian coordinate system on the two-surface which can be imported back into the physical space. Extend this into the 3-space by using gaussian normal coordinates, i.e., lapse = 1, shift = 0, and fixing the normal direction to be the `radial' direction. This defines a quasi-cartesian coordinate system in the neighbourhood of the 2-slice. Now define $\xi^i_{(1)} = (1, 0, 0)$ and so on. Similarly one can define three approximate rotational Killing vectors as $\xi^i_{(X)} = (0, -z, +y)$ and so on so as to define the total angular momentum. These six objects transform in exactly the standard way under the action of the Euclidean group on the flat 3-space.
If one has an asymptotically flat spacelike 3-slice and
considers a sequence of `coordinate spheres' which tend to infinity, this
Brown-Lau-York energy-momentum asymptotes to the ADM energy-momentum. 

In an important recent article Shi and Tam
\cite{ST} showed that the Brown-York energy was positive if the
three-slice was `riemannian', i.e., if $^{(3)}R$, the three-scalar-curvature of some regular filling of $^{(2)}B$,
was positive. However, they could not prove positivity in the general case.
This should come as no surprise as the BY energy may be negative, even in flat
spacetime! I will return to this point.

Motivated by Shi and Tam, Liu and Yau \cite{LY} suggested a new object,
their own quasi-local mass. This is a function of a Riemannian two-surface embedded
in a four-manifold. There is no mention of any 3-surface. It is a function of the expansions along the future and past outer null normals, $\vec{l}, \vec{m}$, of the 2-surface, call them $\rho$ and $\mu$. The lengths of the null normals cannot be specified, but the mutual relationship between the two can. I specify this by demanding $\vec{l}
\cdot \vec{m} = 2$. We are still free to arbitrarily mutually scale each of them, but the product $\rho\mu$ is fixed. The Liu-Yau mass is  
\begin{equation}
8\pi M_{LY} = \int (\sqrt{\rho_0\mu_0} - \sqrt{\rho\mu})dA 
= \int (k_0 - \sqrt{8\rho\mu})dA .\label{LY}
\end{equation}
$\rho_0$ and
$\mu_0$ are the equivalent expansions of an isometric two-surface
embedded in a flat three-slice of flat spacetime. Again one requires the Gauss curvature  of the 2-surface to be positive and also $\rho\mu \ge 0$. This object was introduced earlier by Kijowski \cite{K}. However, the great contribution of Liu and Yau is that they were able to
show that this quantity was always positive if the four-manifold satisfied the
Einstein equations (with well-behaved sources).

Unfortunately, the Liu-Yau mass has significant difficulties. In
particular, it can be shown that surfaces exist in Minkowski space for
which the Liu-Yau mass is non-zero \cite{OMST}. Further, in any
asymptotically flat spacetime, `nice' surfaces exist on which the Liu-Yau mass is unboundedly
large relative to the ADM mass of that spacetime. Alternatives to the Liu-Yau mass have been recently proposed \cite{WY},  with other interesting properties,  but in this letter I will stay with the original definition.

If we have a two-slice embedded in a three-slice we have the following
expressions for the null expansions: $ \rho =  (k +
^{(2)}tr^{(3)}K)$, $\mu = (k - ^{(2)}tr^{(3)}K)$, where $K$ is
the three-extrinsic curvature and $^{(2)}tr^{(3)}K = g_{ab}K^{ab} - n_an_bK^{ab}$, the 2-trace of the 3-extrinsic curvature. These formulae just express the fact that
a null direction is equivalent to one space step plus one time step.
Therefore $\sqrt{\rho\mu} = \sqrt{k^2 - (^{(2)}tr^{(3)}K)^2} < k$. Hence
$k_0 - k < k_0 - \sqrt{\rho\mu}$ and so the Liu-Yau mass is bigger
than the Brown-York energy, except when  $^{(2)}tr^{(3)}K \equiv 0$ on the entire  2-surface.

If we have a two-surface in a four-manifold we have a well-defined Liu-Yau
mass. The Brown-York energy is not well-defined, it depends on the choice of
three-slice in which we embed the two-surface. In fact we can show that the
maximum value of the Brown-York energy equals the Liu-Yau mass, at least if $\rho\mu >0$. In other
words,  given any two-surface, there exists a local three-surface in which it
is embedded and the Brown-York energy relative to this slice equals the
Liu-Yau mass. The
definitions of the null expansions and their normalizations are such that one
can perform a relative scaling of the null vectors so that the two null
expansions become equal. The difference between the two null
vectors defines a future-pointing time direction. The null vectors on the
surface cannot have a twist, otherwise they will not be surface-forming.
This guarantees that their difference will also be twistfree. Thus
this timelike vector generates a local spacelike three-slice orthogonal
to it on which the quantity $(^{(2)}tr^{(3)}K)$ vanishes on the
chosen two-surface and so the Liu-Yau mass of the given two-surface equals the
Brown-York energy as defined for this special slice.

In Physics, one must distinguish between the energy and the mass. The
mass is a frame-independent quantity while the energy is the zeroth component
of a four-vector and it undergoes a Lorentz transformation under a boost. We
expect a `mass' to be smaller than an `energy'. Therefore this relationship
between the Liu-Yau mass and the Brown-York energy should make us stop and
worry. In any asymptotically flat spacetime there exist natural 
asymptotically flat spacelike slices. These are the ones on which the ADM
energy-momentum is well defined. Choose a large `round' two-sphere on such a
slice. More-or-less any reasonable definition of `round'  can be used. We know that the Brown-York energy approaches the ADM energy and the
Brown-Lau-York momentum approaches the ADM momentum on such spheres.  By 
choosing a suitably boosted slice we can make the Brown-York energy as
large as we please, and therefore push up the Liu-Yau mass.

On such a boosted slice, with such a two-surface, we have
\begin{equation}
M_{LY} \ge E_{BY} \approx E_{ADM} = \gamma M_{ADM}
\end{equation}
where $\gamma$ is the boost parameter.  Therefore we can make the ratio $M_{LY}/M_{ADM}$ as large as we please.

The fact that the Liu-Yau mass can be made unboundedly large holds true even
in Minkowski space. Consider one of these nontrivial surfaces in Minkowski
space on which the Liu-Yau mass is nonzero \cite{OMST}. Define the surface with respect to
some interior point in terms of two angles, $(\theta, \phi)$, by giving two
functions
$r = R(\theta, \phi)$ and $t = T(\theta, \phi)$, where $(r, t)$ are the standard
flat space-time coordinates. Now scale up this surface by multiplying the two
functions by some large constant
$C$, i.e.,
$(R, T) \to (CR, CT)$. The mean curvatures will scale like $1/C$ but the area
scales like
$C^2$. Therefore the Liu-Yau mass scales like $C$. This should not be
surprising. We know that Minkowski space has no intrinsic mass or length
scale. Therefore this kind of mass scaling must always be possible.

The Brown-York energy can also be made unboundedly large by choosing a
sufficiently large boost. This, however, is not a major drawback because the
Brown-Lau-York momentum also becomes unboundedly large so that the Brown-York
mass remains small and asymptotes to the ADM mass. The lack of a
Liu-Yau momentum can be seen as a key source of our difficulty. Such an object would be very useful. I will return to this issue.

This led me
to consider spherically symmetric spacetimes (and spherical slices thereof).
 Spherical symmetry guarantees that
the linear momentum must vanish, and no meaningful  distinction between `energy' and `mass' can be made. In that special case I find that the
Liu-Yau mass (or energy!) is remarkably interesting.

Let us consider the moment of time symmetry slice of the Schwarzschild
solution and work out the Brown-York energy for a round two-sphere in it.
This equals the Brown-York mass and also the Liu-Yau mass as the
extrinsic
curvature vanishes. We get
\begin{equation}
E_{BY} = M_{BY} = M_{LY}= m\left (1 + \frac{m }{ 2r}\right )
\end{equation}
where $m$ is the Schwarzschild mass and $r$ is the radius of the sphere
in
isotropic coordinates. This means that $r = m/2$ is the throat and
$r = 0$ is the `other' end. The relationship between the Schwarzschild
$R$
and the isotropic $r$ is $R = r(1 + m/2r)^2$. In Schwazschild coordinates
the expression is
\begin{equation}
E_{BY} = M_{BY} = M_{LY}= r\left (1 \pm \sqrt{1 - \frac{2m }{ r}}\right ),
\end{equation}
where the minus sign holds in the right, outer, quadrant of the extended
Schwarzschild solution and the plus sign should be used in the left
quadrant.

Many years ago three of us (Bizon, Malec, and I) investigated the
binding
energy of spherical stars \cite{BMOM}. In particular, we solved the
initial constraints for a thin spherical shell at rest. This is
interesting because this is the configuration of a given size with least
binding energy. We found
\begin{equation}
M = m\left (1 + \frac{m }{ 2r}\right ),
\end{equation}
where $M$ is the mass of the shell, $m$ is the Schwarzschild mass of the
solution, and $r$ is the radius in isotropic coordinates of the shell,
exactly the same as $M_{LY}$! 

Given a spherical spacelike two surface in a spherical space time, there are
many different regular spherical continuations into the interior (regularity
implies some smoothness and  that the matter satisfies the weak energy
condition). Let me consider those configurations where the interior matter is instantaneously at rest.  If I assume a regular center, there must be some nontrivial mass
density,
$\rho$,  inside. I now compute the interior mass content,
$\int \rho dv$. Note that I take the proper three integral
over the given three geometry. It turns out that the Liu-Yau mass (or the maximum of the
Brown-York energy) of the given two-surface equals the minimum of the mass
content over all instantaneously stationary fillings. The minimum is achieved by having a mass
shell just inside the given two surface. 

This is easy to show by using isotropic coordinates. We have a spherically symmetric, moment of time symmetry initial data with some positive density distribution, $\rho(r)$, of compact support. The 3-metric is conformally flat, with conformal factor $\phi$.  The Hamiltonian constraint becomes $\nabla^2 \phi = -2\pi \phi^5\rho$. The conformal factor $\phi = 1 + m/2r$ outside the support of the matter. The divergence theorem then gives $m = \int \phi^5\rho d^3x$, while the matter content is given by $M = \int \rho dv = \int \phi^6 \rho d^3x$. The maximum principle tells us that $\phi$ is monotonically decreasing outwards. Therefore the minimum of $\phi$ in the integral for $M$ occurs on the boundary of the star, say at $r = r_0$.
Therefore we have
\begin{equation}
M \ge \phi_{(min)} \int \phi^5 \rho d^3x = \left (1 + \frac{m}{2r_0}\right ) m.
\end{equation}
This inequality becomes an equality when the matter is concentrated at the minimum of the potential, i.e., at the boundary of the `star'.

There are many quasi-local mass expressions which can be used in spherically
symmetric spacetimes. Let me discuss just two of them. The Misner-Sharp mass
is \cite{MS}
\begin{equation}
M_{MS} = \frac{\sqrt{A} }{ 64\pi^{\frac{3}{2}}} \int (k^2_0 - 8\rho\mu)dA
\label{MS},
\end{equation}
in terms of the same objects used to define the Liu-Yau mass. This is a constant in vacuum and equals the Schwarzschild mass  if there is no
matter outside the given surface. Therefore we can view it as telling us about
possible exteriors but nothing about the possible interiors. Another standard
expression is the Bartnik mass
\cite{RB}. Again, if we have a regular vacuum exterior, the Bartnik mass
equals the Schwarzschild mass, so again it tells us about possible
exteriors, nothing about interiors.

 All of the standard quasi-local mass
definitions in the Schwarzschild solution give the Schwarzschild mass,
independent of the surface. We have grown so used to this as the `correct'
answer that people found the fact that the Brown-York mass (and, by
extension, the Liu-Yau mass) was not constant disturbing. The fact that the
mass increased as one moved inward was even more disturbing. As one moves
inward, one is moving down in the gravitational potential, the gravitational
red-shift is increasing. Any matter inside the chosen surface is deep inside
the potential well and its contribution to the `mass at infinity' is
equivalently diminished. Therefore the smaller the surface, the more matter
we need to generate the the given Schwarzschild mass. Therefore any quantity
which tries to capture the amount of matter inside the given surface has to
increase as the surface shrinks.

Spherical symmetry is a situation where the Liu-Yau mass has a real
advantage over the Brown-York mass. Consider a round two-surface, of radius
$r = R_0$, in Minkowski space. This has zero Liu-Yau mass. Now assume that it
is embedded in a nontrivial spherically symmetric spacelike three slice. This,
locally (at the surface) can be parametrised as $t = C(r - R_0)$ where $(r,
t)$ are standard flat spacetime coordinates and $|C| < 1$. We have $k_0 =
2/R_0$. However, the unit proper radial distance in the slice is $\Delta r =
1/\sqrt{1 - C^2} > 1$. Therefore $k = 2/\sqrt{1 - C^2}R_0 > k_0$. Hence the
Brown-York mass is negative for this slicing. By having $C \approx 1$, one
can make the Brown-York mass as negative as one pleases.

It is clear that the idea that the Liu-Yau mass gives a lower bound on the
enclosed energy cannot continue to hold in a general spacetime. Consider a
surface in a vacuum solution to the Einstein equations. There will be a
vacuum interior, but the Liu-Yau mass will, in general, be non zero. The
energy is supplied by the gravitational waves. These cannot exist in a
spherical spacetime, so the energy has to be supplied entirely by `matter'.
The other situation where we do not expect any gravitational radiation is
when the solution is static (or stationary). It would be nice to find some
relationship between the Liu-Yau mass and interior matter in a general static
spacetime. The existence of surfaces in Minkowski space (the simplest static
solution) with non zero Liu-Yau mass shows that this is a vain hope. The only
possible hope is to restrict attention to two surfaces which lie in the
`static' (moment of time symmetry) slice. 

If we think of the Liu-Yau mass as somehow measuring the interior energy we
should not be too surprised that its value on a two-surface in a
boosted slice be large. Even though there may be no matter inside, the
interior gravitational radiation will partake of the boost and thus the
`gravitational wave energy', however one can define it, will be large. 

It is
very tempting to try and convert the Liu-Yau mass into an energy. This
involves finding a way to adjoin a momentum to it. It is easiest to do this by   choosing a `preferredÕ 3-slice. Given a two-surface, with positive null expansions,  one obvious such
slice is the one in which the null expansions are equal, as
discussed above. Now we can compute the Brown-Lau-York energy-momentum on this 3-slice, with the
Brown-York energy equalling the unchanged Liu-Yau mass, which now is to be
regarded as a `Liu-Yau-Brown-York' energy,  and is guaranteed to be positive. This Brown-Lau-Liu-Yau-York energy-momentum has the right asymptotic
behaviour. Consider a sequence of surfaces in a given asymptotically flat
three slice. The new energy-momentum differs from the Brown-Lau-York
energy-momentum as defined relative to the given three slice. However,
the difference shrinks to zero at large radii so that it still asymptotes to the ADM
energy-momentum. 

There are still outstanding questions. Since we can regard the momentum as living in flat space,  we can compute $P^2$ using the flat metric. While we have
the positivity of the Liu-Yau energy,  and we know that, near infinity, $E^2 - P^2 \approx M^2_{ADM}$, it would be wonderful if one could show that $E^2 > P^2$ held in general.

\end{document}